\documentclass[11pt,preprint]{aastex}
\begin{document}
\title{Emergence of a flux tube through a partially ionised solar atmosphere}
\author{T. D. Arber, M. Haynes and J. E. Leake}
\affil{Centre for Fusion, Space and Astrophysics, Physics Department, University of Warwick, Coventry, CV4 7AL, UK\\ \email{T.D.Arber@warwick.ac.uk}}

\begin{abstract}
For a magnetic flux tube, or indeed any flux, to emerge into the Solar corona from the convection zone it must pass through the partially ionised layers of the lower atmosphere: the photosphere and the chromosphere. In such regions the ion-neutral collisions lead to an increased resistivity for currents flowing across magnetic field lines. This Cowling resistivity can exceed the Spitzer resistivity by orders of magnitude and in 2.5D simulations has been shown to be sufficient to remove all cross field current from emerging flux. Here we extend this modelling into 3D. Once again it is found that the Cowling resistivity removes perpendicular current. However the presence of 3D structure prevents the simple comparison possible in 2.5D simulations. With a fully ionised atmosphere the flux emergence leads to an unphysically low temperature region in the overlying corona, lifting of chromospheric material and the subsequent onset of the Rayleigh-Taylor instability. Including neutrals removes the low temperature region, lifts less chromospheric matter and shows no signs of the Rayleigh-Taylor instability. Simulations of flux emergence therefore should include such a neutral layer in order to obtain the correct perpendicular current, remove the Rayleigh-Taylor instability and get the correct temperature profile. In situations when the temperature is not important, i.e. when no simulated spectral emission is required, a simple model for the neutral layer is demonstrated to adequately reproduce the results of fully consistent simulations.

\end{abstract}

\section{INTRODUCTION}
The emergence of new magnetic flux into the Solar corona is responsible for the formation of active regions. The accepted view is that the emergence of $\Omega$-shaped flux tubes through the photosphere is responsible for the formation of sunspots. It is clear therefore that the movement of magnetic flux from the convection zone up into the corona is one of the most significant drivers determining the structure of the corona. Furthermore the input of new flux into the preformed corona is often the trigger mechanism for dynamic coronal activity such as prominence eruptions, flares and CMEs. Any attempt at a unified model of Solar activity must couple the magnetic field of the convection zone with that of the corona and hence a full understanding of the flux emergence process which connects them is essential.

The problem with studying flux emergence is that it must couple sub-photospheric plasma with coronal plasma. In traversing this region the magnetic field moves from regions which are convectively unstable to convectively stable, through orders of magnitude changes in equilibrium density, and a rapid increase in temperature. The physics of each of these regions is therefore often dominated by different processes and analytical treatment of the whole emergence process is therefore limited. As a result this subject is now largely investigated by numerical simulations. A typical flux tube, assuming such a well defined structure exists in the convection zone, must have sufficient twist to survive its transit of the convection zone \citep{moreno1996,dorch1998}. It will then reach the photosphere where the buoyancy instability becomes active allowing the flux to escape into the corona \citep{matsumoto1992,murray2006}. Once the flux reaches the region of the chromosphere and corona, where the density drops by six orders of magnitude over a couple of megametres in height, the flux tube expands \citep{matsumoto1993,magara2001,fan2001}. Below the photosphere the plasma beta $\beta>>1$ and the twist of the flux tube gives rise to a $\mathbf{j}\times\mathbf{B}$ force which is easily balanced by small changes in the much larger kinetic pressure terms. In the corona the plasma is characterised by $\beta<<1$ and any emerging  $\mathbf{j}\times\mathbf{B}$ force cannot be balanced by kinetic pressure and thus the flux tube will expand rapidly into a configuration in which any residual $\mathbf{j}\times\mathbf{B}$ force is of the order of $\beta$, i.e. the flux expands until the coronal field is force free. The  initial twist in the emerging flux tube affects the emergence process and how close the coronal field is to force free when it first reaches the corona \citep{abbet2003,murray2006}. This assumes that there is no overlying field with which the emerging flux can interact. Often such a field does exist and the interaction of this new flux with existing magnetic structures can lead to complex, dynamic behaviour \citep{archontis2004,galsgaard2005}. The emerging magnetic flux, since it may have a non-zero $\mathbf{j}\times\mathbf{B}$ force, is also capable of lifting chromospheric material up into the corona. As the field expands, reducing the magnetic forces, this heavier material may trigger the Rayleigh-Taylor instability \citep{isobe2005}.

All of the works cited above have studied flux emergence by using the magnetohydrodynamic (MHD) equations appropriate for a fully ionised plasma. They make the further assumption that the parallel and perpendicular resistivities are equal whereas these differ by a factor $\approx 2$ for a fully ionised plasma. This factor of two is routinely ignored in MHD simulations as the resistivity used, for numerical reasons, exceeds the physical value by many orders of magnitude. For simulations of the coronal or convective regions of the Sun's atmosphere this is a perfectly valid approximation. However, the photosphere and chromosphere are not fully ionised plasma due to their low temperature. It is well known that ion-neutral collisions add an effective anisotropic resistivity into the single fluid equations \citep{cowling1957,braginskii1965}. This additional effect, the Cowling resistivity, acts only on perpendicular currents, i.e. those flowing across the magnetic field, and can be many orders of magnitude larger than the parallel Spitzer resistivity in the chromosphere \citep{khodachenko2004}. This dissipation of perpendicular currents by Cowling resistivity has been used to study the damping of MHD waves in the chromosphere \citep{goodman2000,leake2005} and flux emergence in 2.5D \citep{paper1}. In \citet{paper1} it was shown that the Cowling resistivity was sufficient to destroy all perpendicular currents in 2.5D flux emergence simulations. This is significant as it forces emerged flux to be force free as it traverses the chromosphere. The restricted geometry of 2.5D simulations prevents mass flow along the ignoreable direction and forces the emergence to be arcade like rather than bipolar. The aim of this paper is to return to the simulations in \citet{paper1} and study the effects of the partially ionised layers of the solar atmosphere on flux tube emergence in more realistic 3D geometry.

\section{EQUATIONS AND INITIAL CONDITIONS}
\subsection{Equations}
The standard MHD equations, in Lagrangian form, are modified to include the effects of anisotropic current dissipation. For simplicity it is assumed that the atmosphere is composed entirely of hydrogen. The resulting set of equations apply to a single fluid and include the effects of partial ionisation through the neutral fraction $\xi_n=n_n/(n_i+n_n)$ where $n_n$ is the neutral number density and $n_i$ is the ion number density.

\begin{eqnarray}
\frac{D\rho}{Dt} & = & -\rho\nabla.\mathbf{v} \\
\frac{D\mathbf{v}}{Dt} & = & -\frac{1}{\rho}\nabla P 
+ \frac{1}{\rho}\mathbf{j}\wedge\mathbf{B} + \mathbf{g} + \frac{1}{\rho}\nabla.\mathbf{S}\\
\frac{D\mathbf{B}}{Dt} & = & (\mathbf{B}.\nabla)\mathbf{v} 
- \mathbf{B}(\nabla .\mathbf{v})  \nonumber \\
&& - \nabla \wedge (\eta\mathbf{j_{\|}})
- \nabla \wedge (\eta_{c}\mathbf{j_{\bot}}) \label{b2}\\
\frac{D\epsilon}{Dt} & = & -\frac{P}{\rho}\nabla .\mathbf{v}
+ \frac{\eta}{\rho} {j_{\|}}^{2} + \frac{\eta_{c}}{\rho}{j_{\bot}}^{2} \nonumber \label{energy} \\
&&+ \frac{\varsigma_{ij}S_{ij}}{\rho}-\frac{\epsilon-\epsilon_{0}(\rho)}{\tau}
\end{eqnarray}

Where the parallel and perpendicular current vectors, $\mathbf{j_{\|}}$ and $\mathbf{j_{\bot}}$ respectively, are defined as
\begin{eqnarray}
\mathbf{j_{\|}} & = & \frac{(\mathbf{j}.\mathbf{B})\mathbf{B}}{{|\mathbf{B}|}^{2}} \\
\mathbf{j_{\bot}} & = & \frac{\mathbf{B}\wedge(\mathbf{j}\wedge\mathbf{B})}
{{|\mathbf{B}|}^{2}}
\end{eqnarray}
and $\rho$ is the mass density, $P$ is the gas pressure, $\epsilon$ is the internal specific energy density, $\mathbf{v}$ 
is the centre of mass velocity of the fluid, $\mathbf{B}$ the magnetic field and $\mathbf{g}$ is gravitational acceleration. 
$\mathbf{S}$ is the stress tensor which has components $S_{ij}=\nu(\varsigma_{ij}-\frac{1}{3}\delta_{ij}\nabla.\mathbf{v})$, and
$\varsigma_{ij}=\frac{1}{2}(\frac{\partial v_{i}}{\partial x_{j}}+\frac{\partial v_{j}}{\partial x_{i}}).$

Since the plasma is not fully ionised the total pressure P and specific internal energy $\epsilon$ include the neutral fraction $\xi_n$ through
\begin{equation} \label{pressure}
P=\frac{\rho k_B T}{\mu_m}
\end{equation}
and
\begin{equation} \label{eos}
\epsilon=\frac{k_B T}{\mu_m(\gamma-1)}+\left( 1-\xi_n\right) \frac{X_i}{m_i}
\end{equation}
where $k_B$ is Boltzmann's constant, $\gamma$ is the ratio of specific heats, $\mu_m=m_i/(2-\xi_n)$ is the reduced mass and $X_i$ is the ionisation energy of hydrogen.  Equation \ref{energy} can be used to numerically advance $\epsilon$ but $\xi_n$ is a function of temperature $T$ so Equation \ref{eos} must be solved implicitly for $T$ which can then be used to specify $P$ through Equation \ref{pressure}. For direct comparison with \citet{paper1} and all previous 3D flux emergence simulations here we simply set $\mu_m=m_i/2$ and ignore the $X_i$ term in Equation (8). As discussed in \citet{paper1}
this is unlikely to affect the emergence process through the chromosphere. The chromosphere is not in LTE and the radiation temperature and thermodynamic temperature  cannot be assumed to be the same. The complete calculation of the ionisation state of hydrogen therefore requires the solution of the 3D radiative transfer and ionisation equations. To save time for the 3D emergence problem a simplified reduced model for $\xi_n$ is used. This is based on a modified Saha equation \citep{brown1973} which can be solved for the steady state ionisation equation based on the local temperature with the radiation temperature fixed at the photospheric value \citep{thomas1961}. 
\begin{eqnarray}
\frac{{n_{i}}^2}{n_{n}} & = & \frac{f(T)}{b(T)} \\
f(T) & = & \frac{(2\pi
  m_{e}k_BT)^{\frac{3}{2}}}{h^{3}}\exp\left({-\frac{X_{i}}{k_BT}}\right) \label{ion1}\\
b(T) & = & \frac{T}{wT_{R}}\exp \left[\frac{X_{i}}{4k_BT}\left(\frac{T}{T_{R}}-1
\right) \right]
\end{eqnarray}
where $T_{R}$ is the temperature of the photospheric radiation field 
and $w=0.5$ is its dilution factor.
Using this equation, the ratio of the number density of neutrals to ions
is given by
\begin{equation}
r = \frac{n_{n}}{n_{i}} = 
\frac{1}{2}\left(-1+\sqrt{\left(1+\frac{4\rho/m_{i}}{n_{i}^2/n_{n}}\right)}\right)
\end{equation}
and the neutral fraction $\xi_{n} = \rho_{n}/\rho$ is 
\begin{equation}
\xi_{n} = \frac{r}{1+r}
\end{equation}

The dominant effect of the neutral atoms is how they modify Ohm's law and consequently lead to an anisotropic dissipation of current. The full derivation of the resistive terms in Equations (3) and (4) \citep{cowling1957,braginskii1965} shows that it is the collisions between ions and neutrals which are most important.  For a hydrogen plasma, and in the limit when classical resistivity $\eta$ can be ignored, the Cowling resistivity is given by
\begin{equation}
 \eta_{c} =\frac{{\xi_{n}}^2 B^2}{\alpha_{n}} 
\end{equation}
with
\begin{equation}
\alpha_{n}=\frac{1}{2}\xi_{n}(1-\xi_{n})\frac{\rho^{2}}{m_{n}}
\sqrt{\frac{16k_{B}T}{\pi 
m_{i}}}\Sigma_{in}.
\end{equation}
and $\Sigma_{in} = 5\times 10^{-19} \textrm{m}^{2}$ is the ion-neutral collision cross-section. When these formula are applied to model chromospheres \citep{khodachenko2004} it is found that the Cowling resistivity $\eta_c$ can exceed the classical resistivity $\eta$ by many orders of magnitude. From 2.5D simulations \citep{paper1} it has been shown that the equations above include the dominant corrections to Ohm's law and that the neglect of the Hall term is valid for these flux emergence studies. Furthermore it was also shown that in the upper chromosphere the dominant contribution to Ohm's law is through the Cowling resistivity, not the advective term.

The final term in Equation \ref{energy} is designed to model all of the missing acoustic shock heating, radiative transport and thermal conduction terms. These terms would act to restore the equilibrium photosphere and chromosphere but are too computationally expensive, or simply unknown, and cannot be explicity included. As a first attempt at modelling the detailed heating and cooling terms omitted from Equation \ref{energy} we add a Newton cooling term $-(\epsilon-\epsilon_0)/\tau$ where $\tau$ is the time-scale of the relaxation. The equilibrium specific energy density $\epsilon_{0}$ is chosen to be a function of the density $\rho$.  The reasoning for this is related to the nature of these simulations. The buoyancy force drives magnetic field in the convection zone upwards into the photosphere, where the  field then expands into the atmosphere above. Thus as a parcel of plasma from the convection zone of density $\rho$ is moved upwards into the photosphere, its temperature should be relaxed to its own initial temperature, rather than the local plasma temperature, which is of different density.

A form for the time-scale of this relaxation is required. For this the approach of \citet{gudiksen2005} is adopted. In simulating coronal heating they 
chose $\tau$ to depend on some power of the density
\begin{equation}
\tau = 0.1\left(\frac{\rho}{\rho_{ph}}\right)^{-1.7}
\end{equation}
so that at the relatively dense photosphere ($\rho = \rho_{ph}$) the time-scale is about 0.1s and is large enough that the effect becomes negligible in the sparse corona. 

\subsection{Initial Conditions}

The modified MHD equations are normalised by division of the SI variables
by photospheric values. The basic units are
\begin{eqnarray*}
L_{ph} & = & 150 \, \mathrm{km} \\
v_{ph} & = & 6.5 \, \mathrm{km/s} \\ 
\rho_{ph} & = & 2.7 \times 10^{-4} \, \mathrm{kg/m^{3}}  \\
B_{ph} & = & 1200 \, \mathrm{G}
\end{eqnarray*}
which gives the derived units
\begin{eqnarray*}
t_{ph} & = & 23 \, \mathrm{s} \\
T_{ph} & = & 6420 \, \mathrm{K} \\
P_{ph} & = & 1.2 \times 10^{4} \, \mathrm{Pa} 
\end{eqnarray*}

From here on, unless stated, all quoted values are internal code 
variables and should be multiplied by the above values to recover the SI variables.
The differential equations (1)-(5) are advanced in time numerically using the Lagrangian remap code \texttt{Lare3d} \citep{arber2001}.  
The physical domain simulated extends vertically from -20 (3,000 km below the surface) to 
130 (19,500 km above). The horizontal extend is 75 (11,250 km) about the centre of the domain, i.e.
$-75 \ge x \ge 75$ and  $-75 \ge y \ge 75$.
The $z$-axis is vertical, $y$ across the tube and $x$ aligned with the initial tube axis. 
Simulations have been run on $128^3$, $320^3$ and $512^3$ grids to check convergence. The
computational grid was always uniform so that the minimum grid spacing used was $\Delta x=0.3$,
i.e. $\simeq 44$ km.

The anisotropy in the resistivity prevents the induction equation from being cast in simple diffusive form. 
In order to estimate the relative magnitudes of the resistive and the implicit numerical diffusion contributions
to Equation (3) we consider the 1D model equation $\partial_t B+v \partial_x B = \eta \partial_{xx} B$. 
For the second order accurate scheme employed here the leading order error term introduced in this model
equation is of order $v \Delta x^2 \partial_{xxx} B$ so that if a typical scale length in the dynamic evolution is
$L$ this gives an effective numerical resitivity of $v \Delta x^2 / L$. In the upper chromosphere, where the Cowling
resistivity is dominant, in nomalised units the maximum Alfv\'en speed is 0.6. The worst case for numerical
resolution corresponds to $L\simeq 1$ as this would place three grid points across a slightly diffuse shock. This gives
a normalised implicit numerical resistivity of about $0.06$.  Note that this estimate is based on the fastest phase speed 
and shortest gradient scalelength and thus a value of 0.06 represents an absolute upper limit on the 
numerical resistivity. The typical normalised $\eta_c$ found in the simulations is of order 10, corresponding to a real
magnetic diffusivity of $9.75 \times 10^9 \mathrm{m^2/s}$, and is therefore larger than numerical resistivity. Note that this 
value of diffusivity corresponds to a magnetic Lundquist number of order 0.1. This is larger than that found for estimates
based on the quite chromosphere in \citet{paper1}. This is because chromospheric material expands
as it is lifted due to flux emergence and the associated adiabatic cooling increases the Cowling resistivity.

The initial stratification is a simple 1D model of the temperature profile  of the Sun, 
which includes the upper 3,000 km of the convection zone, the photosphere/chromosphere, 
the transition region, and the base of the corona.
The temperature profile consists of a linear polytrope for the convection zone with a 
vertical gradient at the critical adiabatic value 
\begin{equation}
\frac{dT}{dz} = \frac{\gamma -1}{\gamma}\frac{T}{P}\frac{dP}{dz}.
\end{equation}
The temperature in the photosphere and 
chromosphere is assumed to be constant at 1, 
as is the temperature in the corona at a temperature of 150. These 
two regions are connected by a transition region of width $w_{tr}=5$.
\begin{eqnarray}
T(z) & = & T_{ph} - \frac{g}{m+1}z,\, z<0 \\
     & = & T_{ph} + \frac{(t_{cor}-t_{ph})}{2}\nonumber\\ &&
\left[\tanh \left(\frac{z-z_{cor}}{w_{tr}}\right)+1\right],\, z>0
\end{eqnarray}
$m=\frac{1}{\gamma-1}$ is the adiabatic index for a polytrope, $z_{cor}=25$ 
is the height of the 
corona, $t_{ph}$ is the photospheric temperature and $t_{cor}=150$.  

The density and pressure of the background atmosphere are found from solving the hydrostatic equation 
\begin{equation}
\frac{dP}{dy} = -\rho g
\end{equation}

A magnetic tube is placed in the convection zone at $z=-10$ with the profile
\begin{eqnarray}
B_{x}&=&B_{0}\exp \left(-\frac{r^2}{a^{2}}\right) \\
B_{\phi}&=&qrB_{x}
\end{eqnarray}
where $r$ is the radial distance from the tube centre in the $y,z$ plane. The strength of the field at the centre of the tube, $B_{0}$, is 5 and the the radius of the tube, $a$, is chosen to be 2. $q$ is the amount of twist in the loop defined as
\begin{equation}
q=\frac{B_{x}}{rB_{\phi}}
\end{equation}

This is set to be the minimum required to avoid fragmentation during the rise through the convection zone and is defined as $|q|=1/a$ \citep{moreno1996}.

A choice must be made as to how to initialise the rise of the flux tube in the convection zone. It is thought that flux tubes formed from the toroidal field in the tachocline remain connected to the large scale field by their roots \citep{zwann1978}, while the apex of the tube rises to the surface. As a result a flux tube which reaches the surface will be significantly 'bent' into an $\Omega$-shape. In order to force the tube into this shape in these simulations, the centre is made buoyant while the ends are left in mechanical equilibrium.  This is done by setting the pressure in the tube different to the field-free atmosphere ($p_{0}(z)$) by $p_{1}(r)$ where
\begin{equation}
\frac{d p_{1}(r)}{dr}\hat{\mathbf{e_{r}}} = \mathbf{j}\wedge\mathbf{B},
\end{equation}
so that the pressure gradient matches the Lorentz force. The density in the tube differs from the field-free density ($\rho_{0}(z)$) by $\rho_{1}(r)$ where
\begin{equation}
\rho_{1}(r) = \alpha \frac{p_{1}}{p_{0}(z)}\rho_{0}(z)\exp{\left(-\frac{x^{2}}{\lambda^{2}}\right)}
\end{equation}
where $\alpha$ is used to scale the initial perturbation. In this paper, unless stated otherwise, $\alpha=0.1$.
With this perturbation the centre of the tube, at $x=0$, is buoyant while for $x>\lambda$ the tube is in mechanical equilibrium ($\rho_{1}=0$). The value of $\lambda$ is chosen to be 20, as in \citet{fan2001}.

\subsection{Resistivity Models}
Three different models for the resistivity used in the flux emergence have been studied. The first was the fully ionised plasma model (labelled as FIP in later figures) in which ideal MHD was used. This is the same model used by all previous 3D flux emergence simulations and provides a benchmark against which the effects of Cowling resistivity can be measured. The second is the partially ionised plasma model (labelled as PIP in later figures) which solves the equations including partial ionisation, the Cowling resistivity and the Newton cooling as outlined above. This is the same model as used in \citet{paper1}. The final model is based on a simple model for partial ionisation effects in which a time independent perpendicular resistivity profile is fixed in a layer of the upper chromosphere. This model (labelled as {\em Layer} in later figures) has $\eta=0$ and the Cowling resistivity fixed by
\begin{equation}
\eta_c=400 B^2 \exp[{-(z-10)^2/5}]
\label{layermodel}
\end{equation}
in normalised units. This profile was chosen to closely match the resistivity observed in the chromospheric layers during simulations using the PIP model but still retains the magnetic field dependence.

\section{RESULTS}

\begin{figure}
\plotone{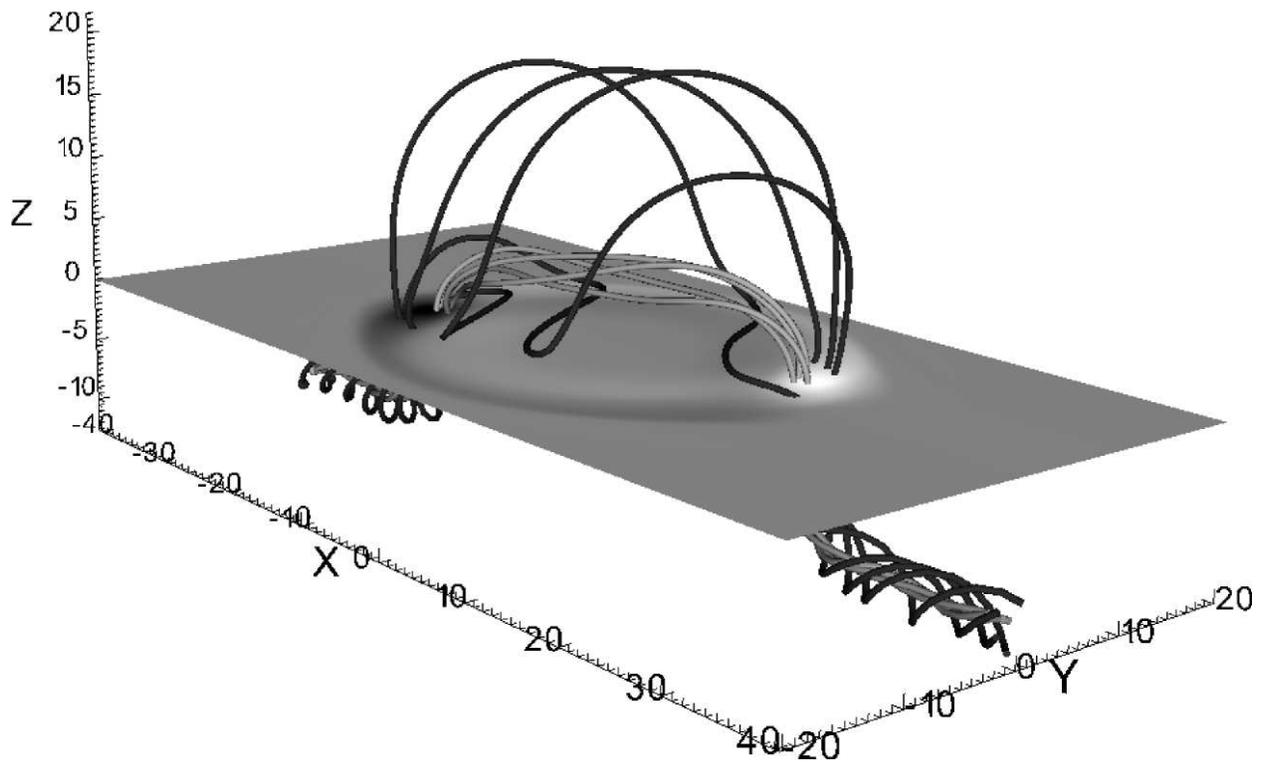}
\caption{Field lines and flux for the partially ionised simulations at $t=160$. The shaded contour plot, at the photopshere $z=0$, shows vertical flux with positive flux shaded dark and negative flux shaded light. Dark field lines are those which connect to the initial equilibrium flux with $r=2.0$ while those shaded lighter correspond to fieldlines near the tube axis, i.e. $r=0.5$.}
\label{fig1}
\end{figure}

\begin{figure}
\plotone{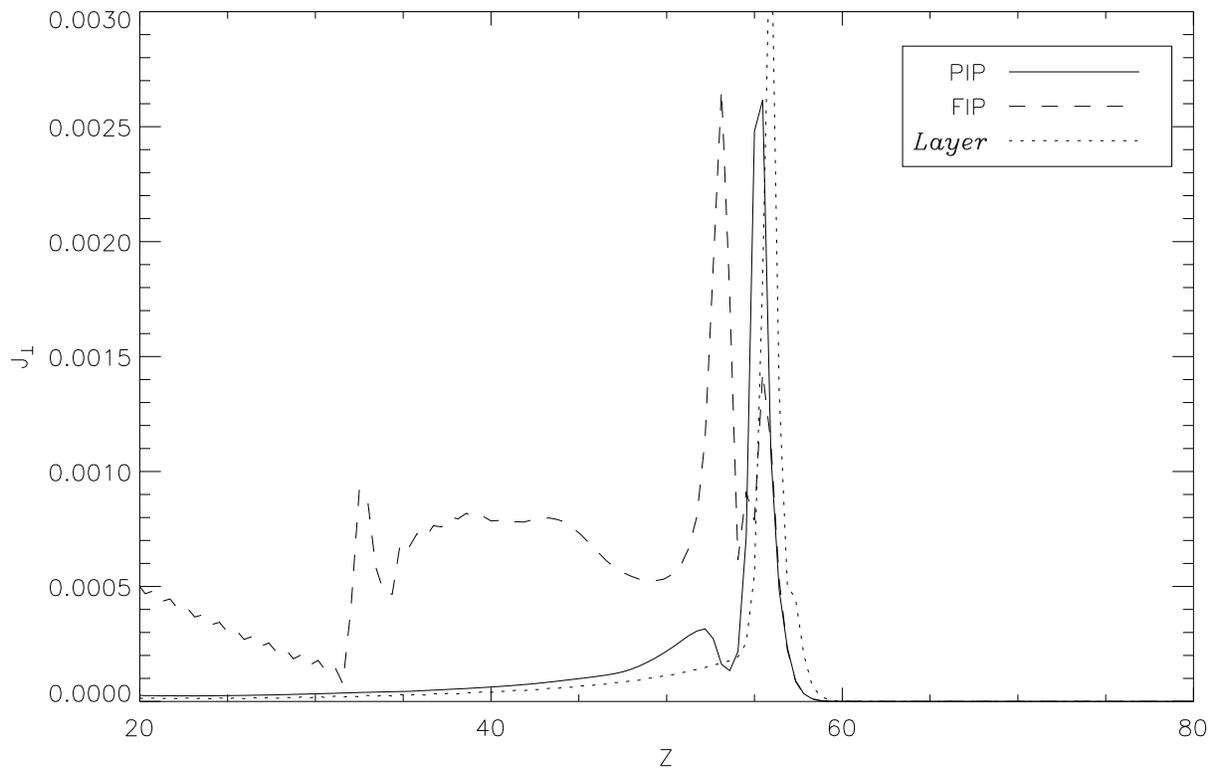}
\caption{Plot of the magnitude of $\mathbf{j_{\bot}}$ as a function of height along the line $x=y=0$ for all three resistivity models at $t=160$.}
\label{fig2}
\end{figure}

The basic stages of the flux emergence process are the same as in previous studies \citep[e.g.][]{fan2001,archontis2004} with the tube initially rising due to buoyancy. When the tube reaches the photosphere the buoyancy stops and the tube expands until sufficient flux has built up for the magnetic buoyancy instability to become important and the field expands through the chromosphere. The structure of the emerged field at $t=160$ can be seen in Figure \ref{fig1}.
In previous 2.5D studies of the effects of the partially ionised layers on flux emergence \citep{paper1} it was possible to quantify these effects by calculating the integrated perpendicular and parallel currents as a function of height. These calculations were at the same time and only included flux inside the expanding envelope of flux. This simple measure of the effectiveness of the Cowling resistivity at removing perpendicular current is not possible in 3D  due to the extra structure discussed below. Figure \ref{fig2} shows the magnitude of $\mathbf{j_{\bot}}$ as a function of height along the line $x=y=0$ for all three resistivity models in the corona. While similar to the results in \citet{paper1} there are a number of significant differences. Firstly the results in Figure \ref{fig2} all show a peak in $\mathbf{j_{\bot}}$ at the top of the emerged flux. This is the expanding shock between flux free and emerging flux regions and was used to define the region {\em inside} which the integrated flux of \citet{paper1} was defined. More significantly the larger $\mathbf{j_{\bot}}$ in the FIP model simulations cannot now be exclusively attributed to $\eta_c$ removing this flux in the PIP simulations. The reason for this is that in 3D the chromospheric material raised into the corona in the FIP
model is sufficient to trigger a Rayleigh-Taylor instability. 

\begin{figure}
\vspace*{-3mm}
\plotone{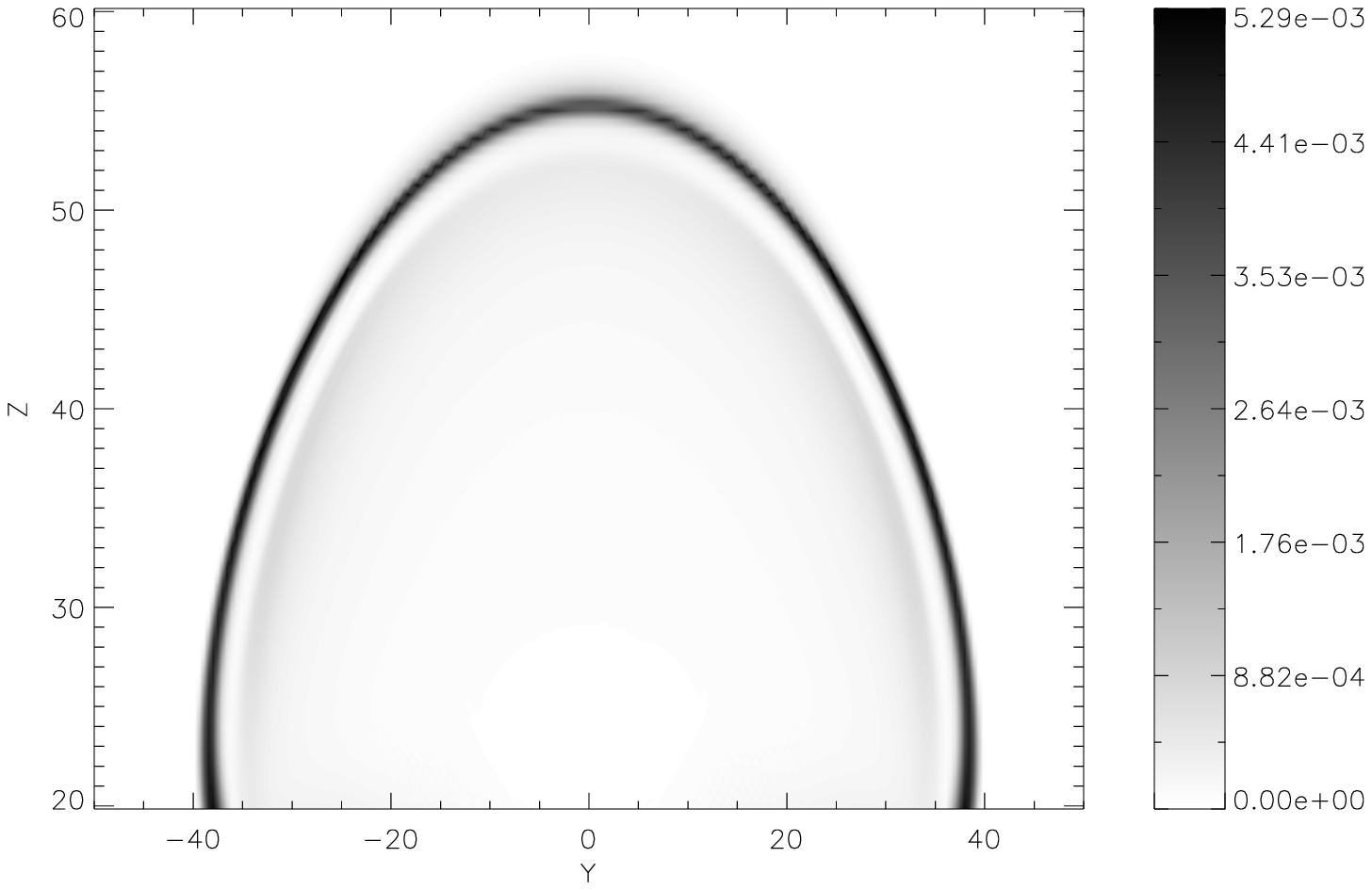}
\plotone{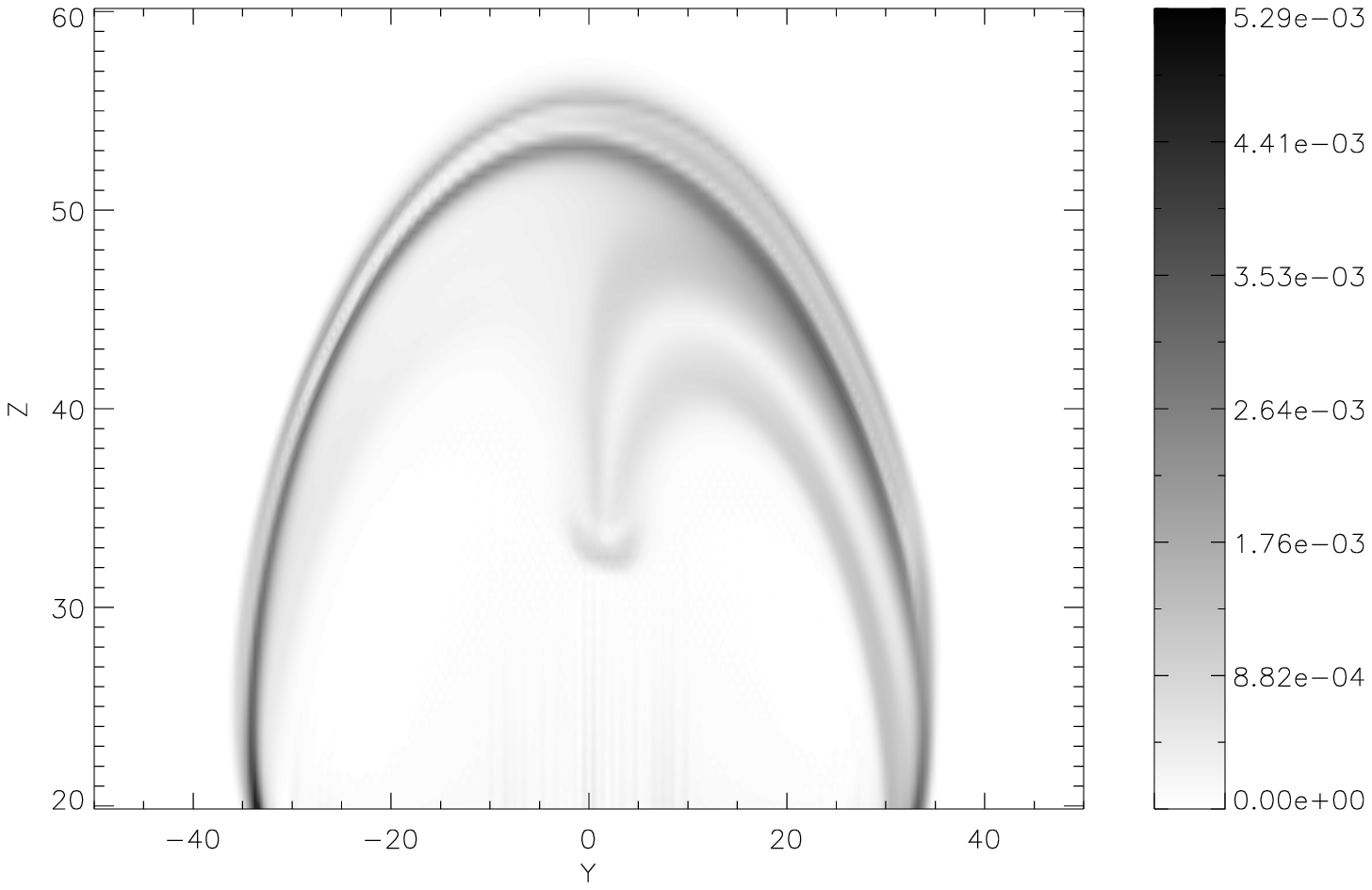}
\caption{Contour plots of the magnitude of $\mathbf{j_{\bot}}$ in a vertical slice through the computational domain in the $x=0$ plane. Results are for PIP model (top) and FIP model (bottom) at $t=160$.}
\label{fig3}
\end{figure}

Figure \ref{fig3} shows the structure of the perpendicular current density in a vertical slice through the centre of the computational domain for the FIP and PIP models. The PIP model shows a slice through a symmetric, expanding shell of flux while for the FIP model there is a dropping central patch of enhanced $\mathbf{j_{\bot}}$. This feature is due to the Rayleigh-Taylor instability, see discussion below, and as this leads to a bending and compression of fieldlines it also contributes to the net $\mathbf{j_{\bot}}$. Hence it is not possible to directly attribute the reduction of $\mathbf{j_{\bot}}$ in the PIP model, compared to the FIP model, shown in Figure \ref{fig2} directly due to the dissipation of $\mathbf{j_{\bot}}$ in the partially ionised chromosphere as some of the excess of $\mathbf{j_{\bot}}$ in the FIP model is created {\em in situ} in the corona by the Rayleigh-Taylor instability.

\begin{figure}
\plotone{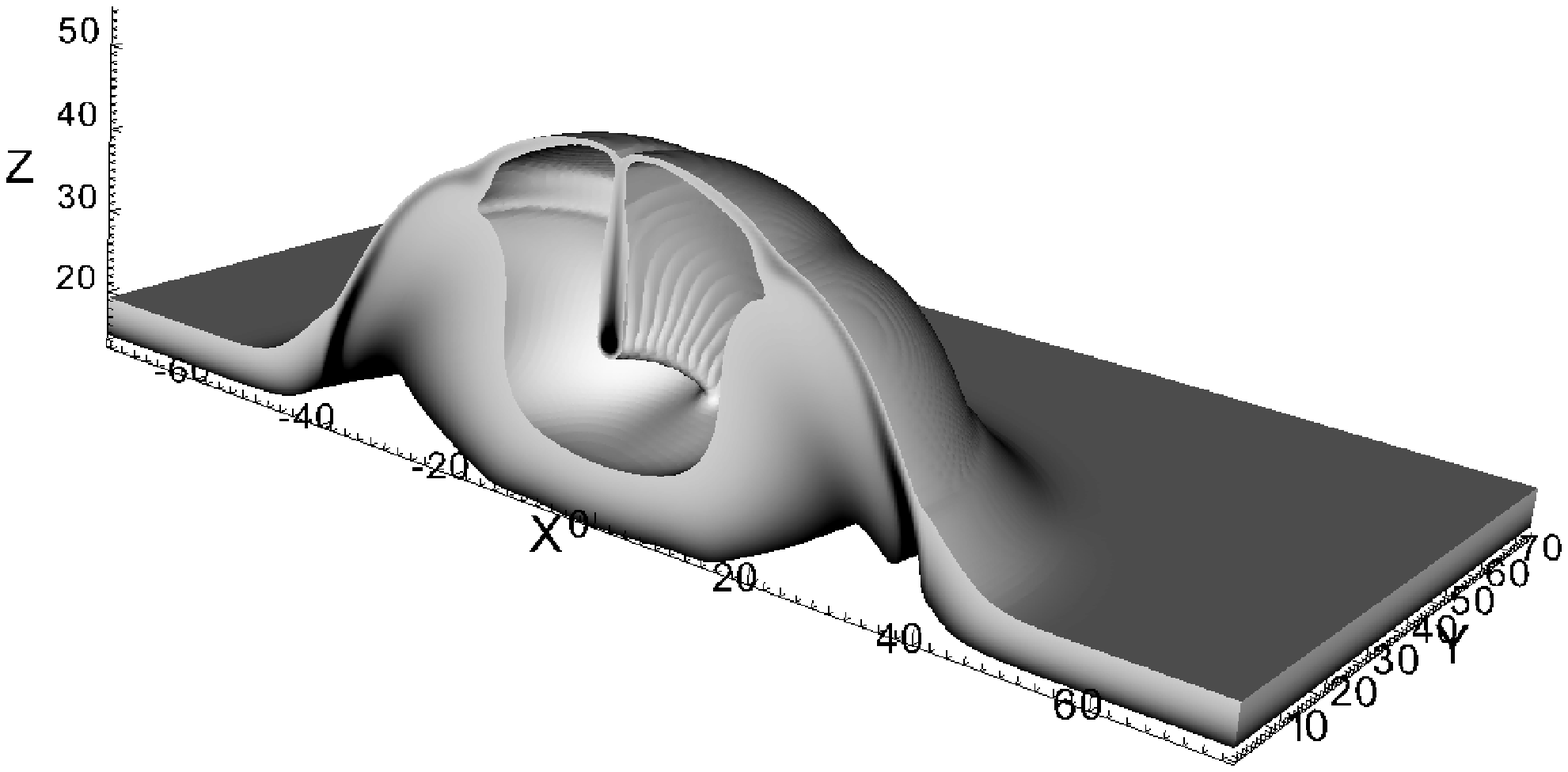}
\caption{Volume rendering of density for the FIP model at $t=160$. The volume contains values of density between $5\times 10^{-7}$ and $5\times 10^{-6}$. The color scaling on the cut through the volume shows higher density as darker shades. The domain has been split along the axis of the initial equilibrium flux tube to show the density structure inside the expanding shell being lifted by the flux emergence. The dominant Rayleigh-Taylor spike can be seen dropping at around $x=0$ in the $y=0$ plane.}
\label{fig4}
\end{figure}

The FIP model allows flux to emerge through the chromosphere with a larger $\mathbf{j \times B}$ force than the PIP model in which the net $\mathbf{j_{\bot}}$  is dissipated by Cowling resistivity. As a result the FIP model lifts more chromospheric material into the corona and is susceptible to the Rayleigh-Taylor instability. This is shown by the isosurface of density in Figure \ref{fig4} with the central density sheet dropping from the Rayleigh-Taylor instability clearly visible. Note that this density sheet is aligned with the magnetic field at the top of the emerging flux and hence, as can be seen from Figure \ref{fig1}, is in a plane which crosses the tube axis. The $x$ direction is therefore not an ignoreable coordinate for this Rayleigh-Taylor mode explaining its absence in the previous 2.5D work.

\begin{figure}
\plotone{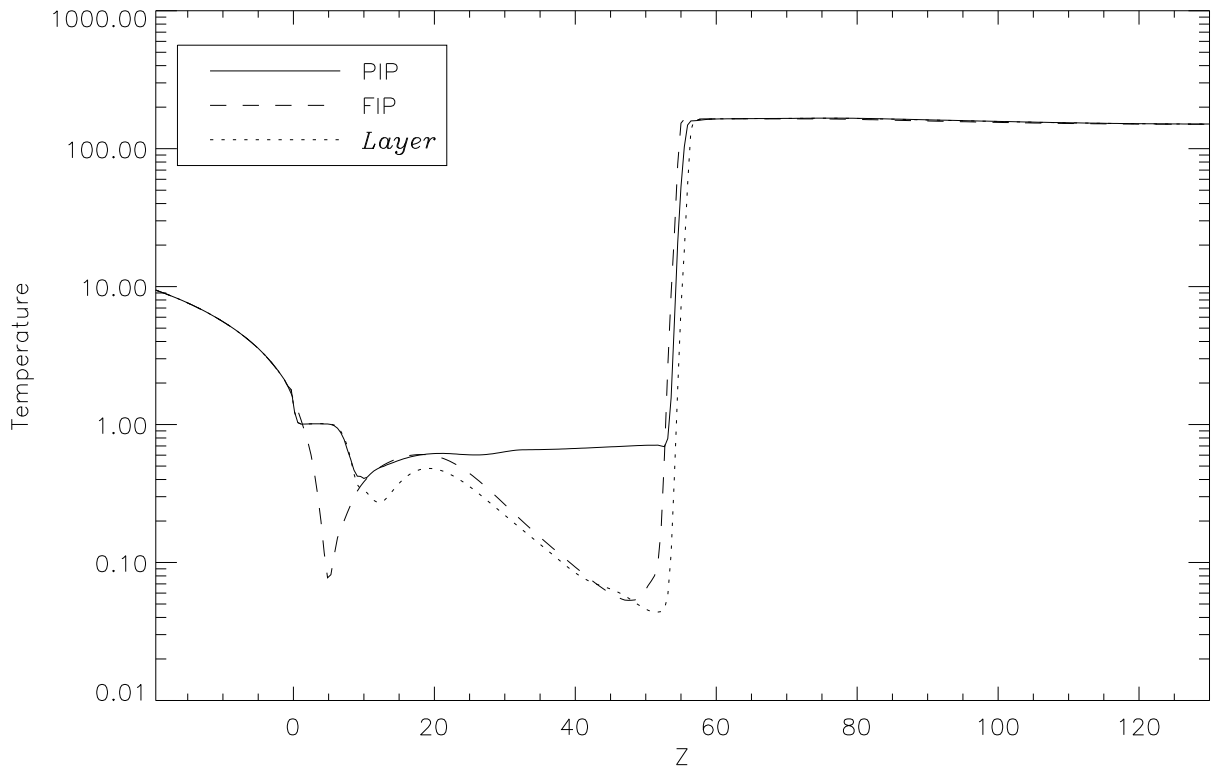}
\caption{Temperature as a function of height for all three resistivity models at $t=160$.}
\label{fig5}
\end{figure}

A common feature of all flux emergence simulations using a fully ionised MHD model is that the rapid expansion of the emerging flux, once it reaches the low density corona, caused adiabatic cooling of the plasma to low temperatures. This can be seen for the FIP model in Figure \ref{fig5} which compares the temperature as a function of height along a line at $x=10$ and $y=0$ where the $x$ coordinate is offset just enough to avoid the Rayleigh-Taylor instability induced density sheet. With the FIP model the temperature in the corona drops to 0.04 in normalised units, or $\simeq 250$K. Including the Newton cooling/heating term with a relaxation time specified in Equation (16) only affects the temperature on the timescale of these simulations in a layer $\simeq 7$ Mm above the photosphere and accounts for the ledge in temperature profile for both the {\em Layer} and PIP simulations between $z=0$ and $z=7$. In the corona only the PIP simulations have a heating term due to the Cowling resistivity and this maintains the temperature there to about 0.7 times the photospheric value.
\section{CONCLUSIONS}

Previous 2.5D simulations \citep{paper1} had shown that including the partially ionised layers of the Solar atmosphere changed the dynamics of magnetic flux emergence by introducing a Cowling resistivity. This dissipated perpendicular current density and lead to a force free coronal field (inside the expanding region of emerged flux but ignoring the perpendicular current density at the interface of emerged flux and field free corona). This simple picture cannot be supported in such a clear way for the full 3D simulations presented here. In 3D the most pronounced difference between partially ionised and fully ionised simulations is that the fully ionised simulations raise more chromospheric material. For the initial conditions used in this paper this meant that the fully ionised model became unstable to the Rayleigh-Taylor instability while the partially ionised model remained symmetric with no signs of instability. The presence of the Rayleigh-Taylor induced perpendicular current density means that it is not possible to assess the affect of the Cowling resistivity on the amount of perpendicular current density emerging into the corona in isolation.

The primary result of these simulations is therefore that the inclusion of the Cowling resistivity affects the amount of chromospheric material uplifted into the corona. The process of removing $\mathbf{j_{\bot}}$ allows most of the field to move through the chomospheric plasma rather than lifting it, thus reducing the amount of mass uplifted.
Recent publications \citep{isobe2005} have suggested that the onset of the Rayleigh-Taylor instability during flux emergence may be the cause of coronal loops. This result may depend critically on the neutral hydrogen but this was absent from all previous flux emergence simulations. 

Often in coronal physics one is only concerned with the emergence of magnetic flux and its structure. In such situations this paper has shown that a greatly simplified model of the Cowling resistivity is capable of reproducing the results of the full PIP simulations, except for the temperature profile. Since the present, and indeed all previous, flux emergence simulations have omitted a full treatment of thermal conduction, radiation effects and coronal heating it is unlikely that such simulations can achieved accurate temprature estimates for emergence.   A practical conclusion from this work is therefore that the minimum physics required to obtain credible field structures in flux emergence is the {\em Layer} model presented in Equation \ref{layermodel}. This has the advantage of being easy to include into any code and yet accurately predicts the correct magnetic field structure and uplifting of chromospheric material. 

\acknowledgements
This work was funded in part by the Particle Physics and Astronomy Research Council. The computational work was support by resources made available through the UK MHD Consortium and Warwick University's Centre for Scientific Computing.

\end{document}